\documentclass{osa-article}

\journal{oe}

\articletype{Research Article}

\begin{document}
\title{A wide-range Prandtl/Schmidt number power spectrum of optical turbulence and its application to oceanic light propagation}

\author{Jin-Ren Yao,\authormark{1,3} Hua-Jun Zhang,\authormark{1} Ruo-Nan Wang,\authormark{1} Jian-Dong Cai,\authormark{1} Yu Zhang\authormark{1,*} and Olga Korotkova\authormark{2}}
\address{\authormark{1}Department of Physics, Harbin Institute of Technology, Harbin, 150001, China\\
\authormark{2}Department of Physics, University of Miami, Coral Gables, Florida 33146, US\\
\authormark{3}jinry@yahoo.com}

\email{\authormark{*}zhangyuhitphy@163.com}

\begin{abstract}
Light influenced by the turbulent ocean can be fully characterized with the help of the power spectrum of the water's refractive index fluctuations, resulting from the combined effect of two scalars, temperature and salinity concentration advected by the velocity field. The Nikishovs' model [Fluid Mech. Res. 27, 8298 (2000)] frequently used in the analysis of light evolution through the turbulent ocean channels is the linear combination of the temperature spectrum, the salinity spectrum and their co-spectrum, each being described by an approximate expression developed by Hill  [J. Fluid Mech. 88, 541562 (1978)] in the first of his four suggested models. The fourth of the Hill's models provides much more precise power spectrum than the first one expressed via a non-linear differential equation that does not have a closed-form solution. We develop an accurate analytic approximation to the fourth Hill's model valid for Prandtl/Schmidt numbers in the interval $[3,3000]$ and use it for the development of a more precise oceanic power spectrum. To illustrate the advantage of our model, we include numerical examples relating to the spherical wave scintillation index evolving in the underwater turbulent channels with different average temperatures, and, hence, different Prandtl numbers for temperature and different Schmidt numbers for salinity. Since our model is valid for a large range of Prandtl number (or/and Schmidt number), it can be readily adjusted to oceanic waters with seasonal or extreme average temperature and/or salinity or any other turbulent fluid with one or several advected quantities.  
\end{abstract}

\section{Introduction}
The growing interest in the development of optical underwater imaging systems \cite{Hou:09}, laser communications \cite{Yi:15,Baykal:18BER,Cui:19} and remote sensing schemes \cite{Korotkova_18Detect} stipulates the request for the accurate predictions of the oceanic turbulence effects on the light wave evolution. Such effects can be characterized with the help of the parabolic equation, the Rytov approximation, the extended Huygens-Fresnel integral and other classic methods as long as the refractive-index power spectrum of turbulent fluctuations is established \cite{Phillips2005Laser, KorBook, PhysRevA.93.033821, PhysRevA.97.043817}. As applied to oceanic  propagation, the theoretical predictions for the major light statistics of the optical waves have been established (see recent review \cite{KOROTKOVA_Lightocean}) on the basis of the widely known power spectrum developed by Nikishovs \cite{Nikishov2000Spectrum}.  In particular, the spectral density and the beam spread for coherent and random beams have been analyzed \cite{Wei2006Influence}, the spectral shifts in random beams were revealed \cite{FarwellOKSpectrum}, the polarimetric changes of the random electromagnetic beams have been discussed \cite{FarwellOKPolarization}, the loss of coherence of initially coherent beams has been addressed \cite{FarwellOKCoherence}, the scintillation analysis was provided in Refs. \cite{FarwellOKScintillation} and \cite{BaykalStrongSI} and the structure functions of various waves were derived in Refs. \cite{BaykalStructure} and \cite{SecondStructure}.  

In the situations when there is more than one scalar field advected by turbulence, the total refractive-index power spectrum can be expressed as a linear combination of the individual spectra and their co-spectra. For instance, in the case of the oceanic turbulence with advected temperature and salinity (sodium chloride) concentration, the total spectrum is the linear combination of three spectra: temperature spectrum, salinity spectrum, and their co-spectrum \cite{Nikishov2000Spectrum} (see also \cite{JinrenYao_17}).  An important dimensionless quantity determining the turbulent spectrum profile is the Prandtl number, $Pr_{\rm{T}}$, defined as a ratio of viscous diffusion rate (kinematic viscosity) $[\rm{m}^2\rm{s}]$ to the thermal diffusion rate (thermal diffusivity) $[\rm{m}^2\rm{s}]$. 
Another important dimensionless quantity is the Schmidt number, $Pr_{\rm{S}}$, defined as the ratio of kinematic viscosity and mass diffusivity. Since the Prandtl numbers for temperature and the Schmidt numbers for salinity in the water are sufficiently large (greater than one), their power spectra have three regimes: \textit{inertial-convective} regime with the Kolmogorov power law $\Phi_n(\kappa) \sim \kappa^{-11/3}$ at the low wave numbers, \textit{viscous-convective} regime with the power law $\Phi_n(\kappa) \sim \kappa^{-3}$ for higher wave numbers and the \textit{viscous-diffusive} regime at even higher wave-numbers where the spectra decrease very rapidly driven by the diffusion of the advected quantity. Due to two different power laws at the high wave numbers each spectrum forms a characteristic ``bump'' \cite{Hill2016Spectra, Hill_Optpro, Yi:18}. 

To achieve a more precise match of the spectra at the boundaries between different regimes, one of the four models proposed by Hill \cite{hill_1978} may be applied. Among them, the Hill's model 1 (H1) that provides the simple analytic approximation but lacks precision in accounting for the bump, and the Hill's model 4 (H4) that is more precise but involves a non-linear differential equation without known analytic solution \cite{ednote_02}.

The widely used Nikishovs' oceanic turbulence spectrum \cite{Nikishov2000Spectrum}  is developed as a linearized polynomial involving temperature spectrum, salinity spectrum and their co-spectrum, each being based on H1 (see Appendix I for more details). Hence, it has a simple analytic form but lacks precision. It can, in principle, be applied to various Prandtl/Schmidt numbers but, as we will illustrate, substantially deviates from H4, especially for Prandtl/Schmidt numbers much larger than one. Also, a recent model for the oceanic power spectrum \cite{Yi:18,Li:19} was obtained by fitting H4 at specific $Pr$ values: $Pr  = 7$ for temperature spectrum, $Pr  = 700$ for salinity spectrum and $Pr  = 13.86$ for their co-spectrum. It was revealed that the oceanic power spectrum of \cite{Yi:18} has a substantially higher bump at high spatial frequencies as compared with that for the Nikishov's spectrum, resulting in a greater scintillation index in light waves interacting with such a medium.

Moreover, within the last two decades, several mathematically tractable models based on H4 have also been introduced for atmospheric refractive-index spectrum \cite{Churnside1990A,Frehlich_1992,Andrews_1992,Grayshan2007,Muschinski_2015TemVa}, by fitting $Pr = 0.72$ and $Pr = 0.68$ for temperature and humidity Prandtl numbers, respectively. For such low Prandtl numbers in the atmospheric turbulence the viscous-convective regime is not that prevalent but nevertheless the high-frequency bump can still be formed. Among the atmospheric power spectrum models are the Frehlich's model being a fifth-order Laguerre expansion \cite{Frehlich_1992}, as well as the Grayshan's and the Andrews' models both being products of polynomials and Gaussian  functions \cite{Grayshan2007,Andrews_1992}. All these power spectra provide convenience for theoretical calculations, suggesting that a good analytical fits do not have to be mathematically involved.

The aim of this paper is to employ the H4 model for developing a power spectrum that can be readily adapted to the wide range of Prandtl number and Schmidt number but, at the same rate, remain numerically accurate and mathematically tractable. Such new spectrum can be applied for classic homogeneous oceanic turbulence in a variety of thermodynamic states. Moreover, it can also be employed for optical characterization of other turbulent media, with one or two advected quantities, for example, fresh turbulent water contaminated by a chemical.  The ratio of kinematic viscosity to diffusivity is orders of magnitude larger for temperature and salinity fluctuations in water than for temperature and humidity fluctuations in air \cite{Hill_Optpro}, resulting in much larger Prandtl numbers and Schmidt number (averaging at $Pr_{\rm T}=7$ for temperature and $Pr_{\rm S}=700$ for salinity \cite{KOROTKOVA_Lightocean,Nikishov2000Spectrum,Hill2016Spectra,Yi:18,Li:19}. More importantly, these numbers can exhibit variation. For example, at 1 Bar pressure (water surface), at fixed salt concentration at $34.9 \rm{ppt}$, as the temperature increases from $0^{o}$C to $30^{o}$C, $Pr_{\rm T}$ decreases from 13.349 to 5.596 and $Pr_{\rm S}$ decreases from 2393.2  to 456.1 (more details in  Appendix II). On the other hand certain oil-like substances can lead to Prandtl numbers up to $10^5$.

The paper is organized as follows. In section \ref{Sec_1a} we introduce an approximate fit to H4 for $Pr\in\left[ {3,3000} \right]$ for handling situations in which a single scalar is advected by turbulence. Section \ref{Sec_1b} provides the validity estimation of the fit by means of the dissipation constraint. Section \ref{Sec_2a} introduces an H4-based power spectrum of oceanic turbulence by applying the fit of Section 2.1 to three spectra: temperature-based, salinity-based and the their co-spectrum. In Section \ref{Sec_2b} we apply the new spectrum model for analytic calculation of the scintillation index of the spherical wave for a variety of Prandtl numbers and Schmidt numbers and illustrate its agreement with numerical calculation based on the H4 model as well as its discrepancy with the result based on the Nikishovs' spectrum. In Section \ref{Sec_3} all the results are briefly summarized.

\section{Analytic approximation of the Hill's model 4 for one advected quantity}
\label{Sec_1}
\subsection{The polynomial-Gaussian fit of the Hill's model 4}
\label{Sec_1a}

Let us begin by developing the three-dimensional (3D) scalar power spectrum of the refractive index fluctuations produced by a single quantity advected by a fluid, as an analytical fit to the H4 model. According to \cite{Muschinski_2015InvHill} the 3D scalar spectrum $\Phi_n(\kappa )$ is related to a universal dimensionless function $g(\kappa \eta)$ by expression
\begin{align}
\Phi_n (\kappa ) = \frac{1}{{4\pi }}\beta \chi {\varepsilon ^{ - \frac{1}{3}}}{\kappa ^{ - \frac{11}{3}}}g(\kappa \eta),
\label{eq1}
\end{align}
where $\beta $ is the Obukhov-Corrsin constant; $\varepsilon $ is the dissipation rate of the kinetic energy $[{{\rm{m}}^2}{{\rm{s}}^{ - 3}}]$; $\eta $ is the Kolmogorov microscale $[\rm{m}]$; $\chi $ is the ensemble-averaged variance dissipation rate. The units of $\chi$ depend on the advected quantity, for instance, for temperature, salinity and refractive index they are ${{\rm{K}}^2}{{\rm{s}}^{ - 1}}$, ${{\rm{g}}^2}{{\rm{s}}^{ - 1}}$ and ${{\rm{s}}^{ - 1}}$, respectively.

According to model H4, $g(\kappa \eta)$ is the solution of the nonlinear ordinary differential equation \cite{hill_1978,Muschinski_2015InvHill}
\begin{align}
\frac{{\rm{d}}}{{{\rm{d}}z}}\left\{ {{{({z^{2b}} + 1)}^{ - \frac{1}{3b}}}\left[ { - \frac{{11}}{{\rm{3}}}f(z){\rm{ + }}z\frac{{{\rm{d}}f(z)}}{{{\rm{d}}z}}} \right]} \right\} = \frac{{22}}{3}c{z^{\frac{1}{3}}}f(z),
\label{eq2}
\end{align}
where
\begin{align}
f(z) = g(az) = g(\kappa \eta ),
\label{eq3}
\end{align}
and $z = \kappa\eta a^{-1}$, $c = \beta {a^{4/3}}Pr^{-1}$. 
Generally, as recommended by Hill \cite{Hill2016Spectra} the following fixed values of parameters $a = 0.072$, $b = 1.9$, and $\beta  = 0.72$ \cite{ednote_03}
will be used throughout the paper. The numerical solution $g(\kappa \eta)$ of Eq. (\ref{eq2}) exhibits an obvious ``bump'' with height ${g_{\max }}$ gradually increasing with growing values of $Pr$, for $Pr  > 0.2$ \cite{Muschinski_2015InvHill}. 

Considering the mathematical convenience given by Gaussian function in related calculation \cite{Yue:19,Phillips2005Laser},
we keep the form --- a product of a polynomials and a Gaussian function \cite{ednote00} --- from Andrews' model \cite{Andrews_1992}.
In addition,
to include the significant effect of $Pr$ on $g(\kappa \eta)$ \cite{Muschinski_2015InvHill}, 
we separate $(\kappa \eta )$ and $c$, and give them different values of power \cite{ednote01}:
\begin{align}
\nonumber g(\kappa \eta ) = &\left[ {1 + {h_1}{{(\kappa \eta )}^{h_2}}{c^{h_3}} + {h_4}{{(\kappa \eta )}^{h_5}}{c^{h_6}}} \right]\\
&\times\exp \left[ { - {h_7}{{(\kappa \eta )}^2}{c^{h_8}}} \right],
\label{eq07_1}
\end{align}
where ${h_2}$, ${h_5}$ and ${h_7}$ are limited to positive numbers and then fitted the numerical solution $\nonumber g(\kappa \eta )$ of Eq. (2) with Eq. (4) by iterative least squares method. This resulted in the following expression for $\nonumber g(\kappa \eta )$:
\begin{align}
	\nonumber g(\kappa \eta ) = &\left[ {1 + {\rm{21}}{\rm{.61}}{{(\kappa \eta )}^{{\rm{0}}{\rm{.61}}}}{c^{{\rm{0}}{\rm{.02}}}} - {\rm{18}}{\rm{.18}}{{(\kappa \eta )}^{0.55}}{c^{{\rm{0}}{\rm{.04}}}}} \right]\\
	&\times\exp \left[ { - 174.90{{(\kappa \eta )}^2}{c^{{\rm{0}}{\rm{.96}}}}} \right].
	\label{eq5}
\end{align}
valid in the interval $Pr\in\left[ {3,3000} \right]$. On substituting from Eq. (\ref{eq5}) into Eq. (\ref{eq1}), we find that the fitted power spectrum based on H4 takes form
\begin{align}
	\nonumber {\Phi}_n(\kappa ) = &\left[ {1 + {\rm{21}}{\rm{.61}}{{(\kappa \eta )}^{{\rm{0}}{\rm{.61}}}}{c}^{{\rm{0}}{\rm{.02}}} - {\rm{18}}{\rm{.18}}{{(\kappa \eta )}^{0.55}}{c}^{{\rm{0}}{\rm{.04}}}} \right]\\
	&\times \frac{1}{{4\pi }}\beta {\varepsilon ^{ - \frac{1}{3}}}{\kappa ^{ - \frac{11}{3}}}{\chi }\exp \left[ { - 174.90{{(\kappa \eta )}^2}{c}^{{\rm{0}}{\rm{.96}}}} \right].
	\label{eq_a1}
\end{align}

\begin{figure*}[!b]
	\centering
	\includegraphics[width=0.48\textwidth]{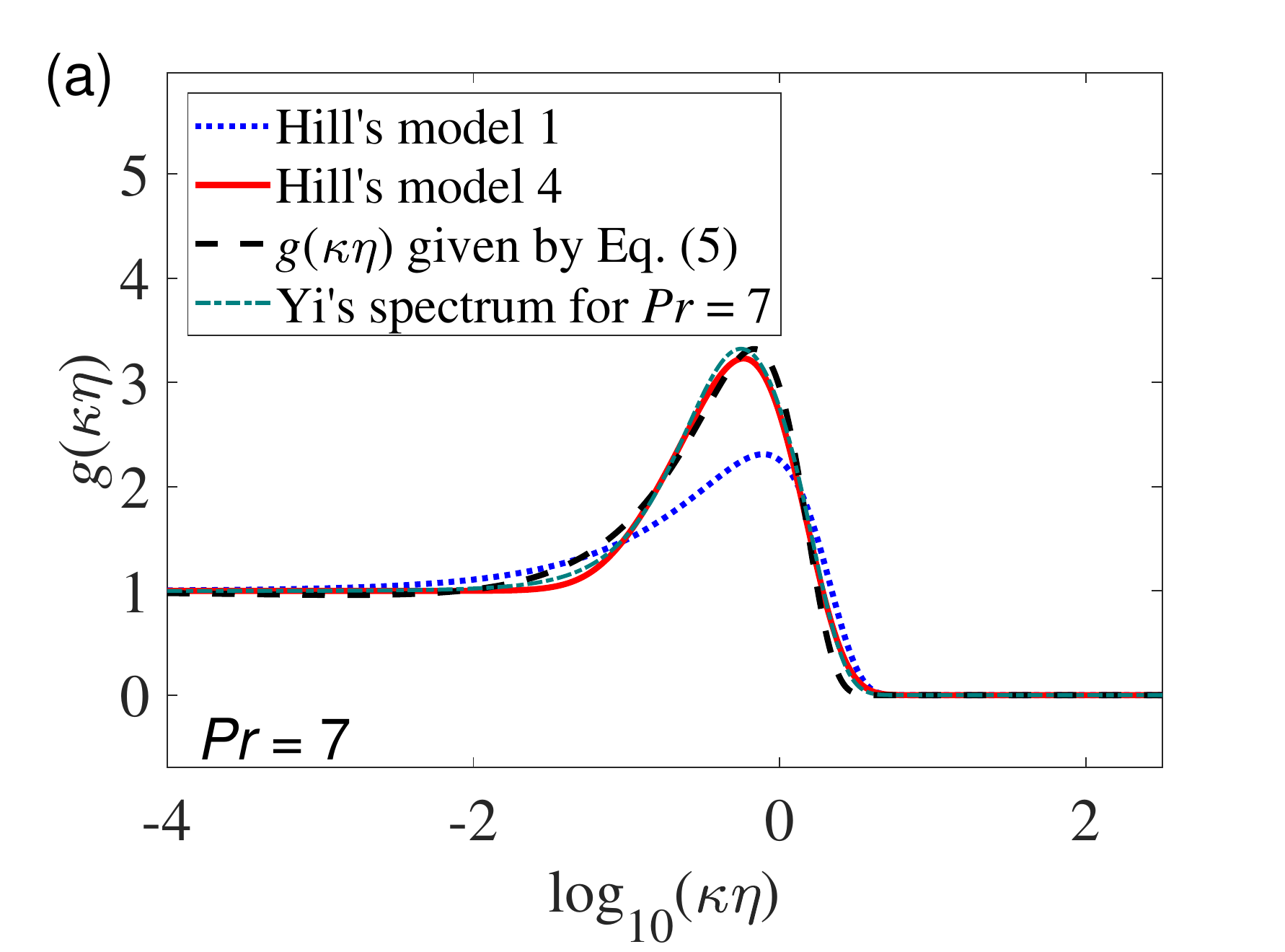}
	\includegraphics[width=0.48\textwidth]{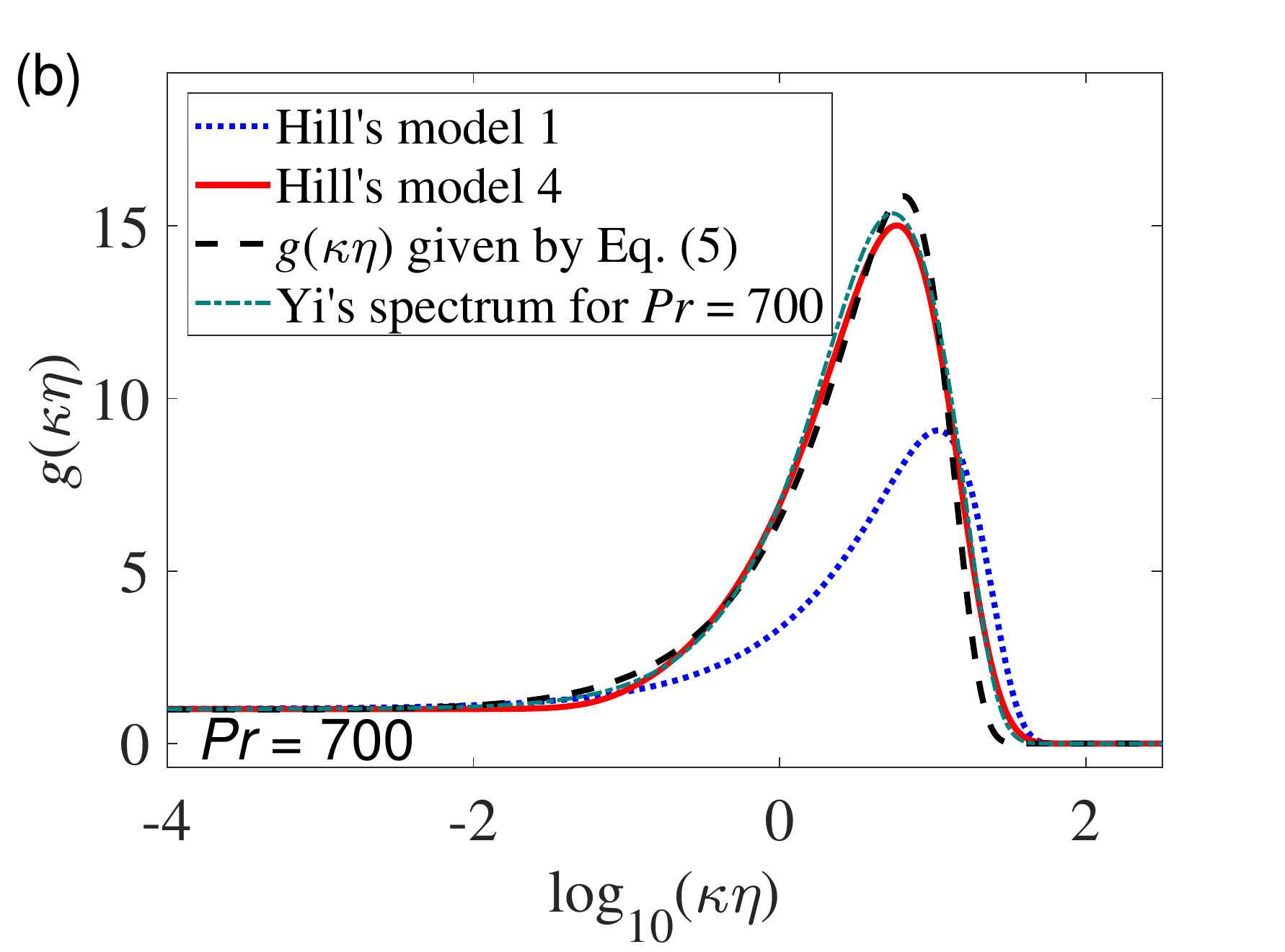}
	\caption{Universal dimensionless function $g(\kappa\eta)$ at different Prandtl numbers: (a) Pr=7, (b) Pr=700.}
	\label{fig1}
\end{figure*}

\begin{figure*}
	\centering
	\includegraphics[width=0.48\textwidth]{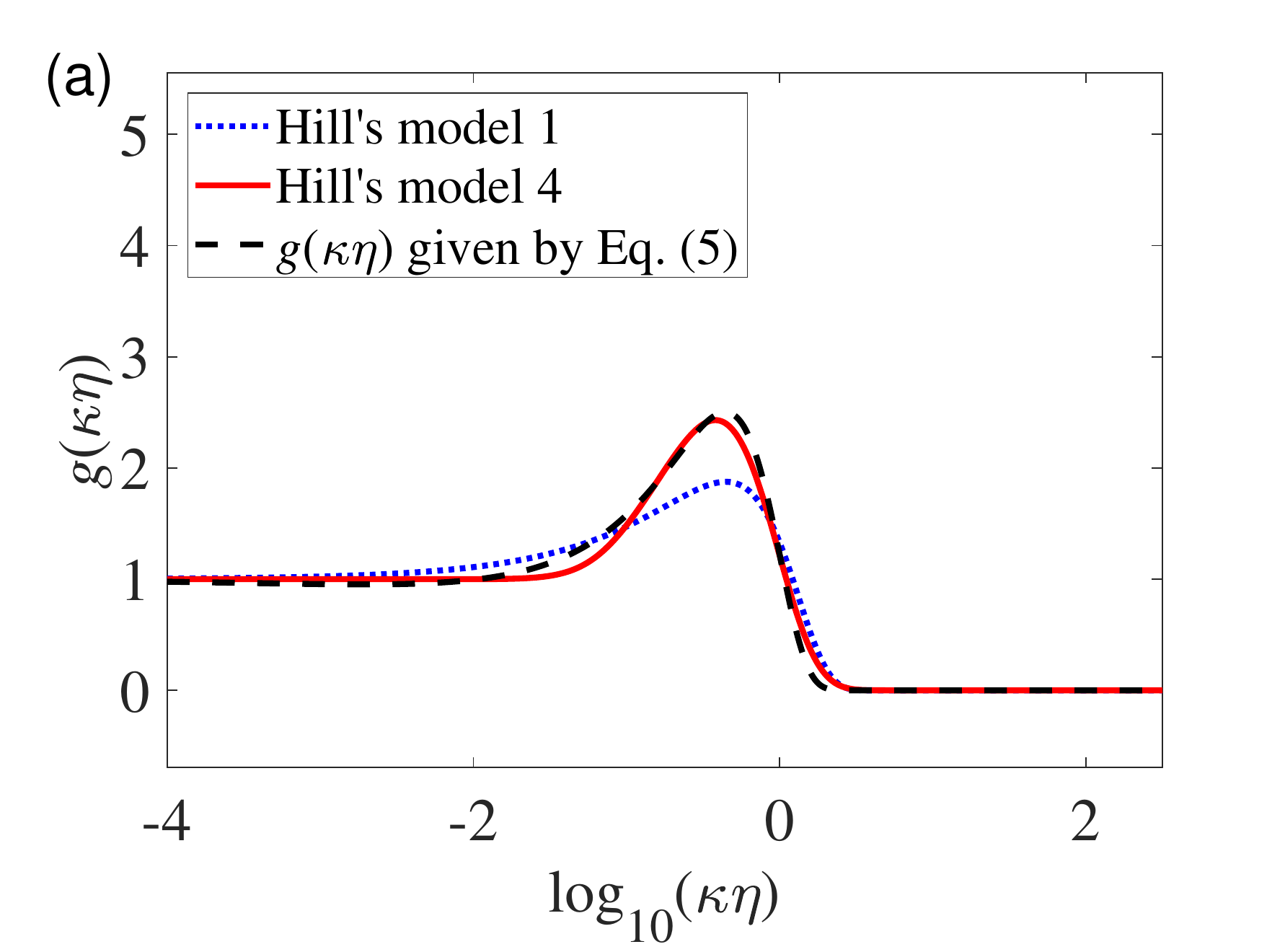}
	\includegraphics[width=0.48\textwidth]{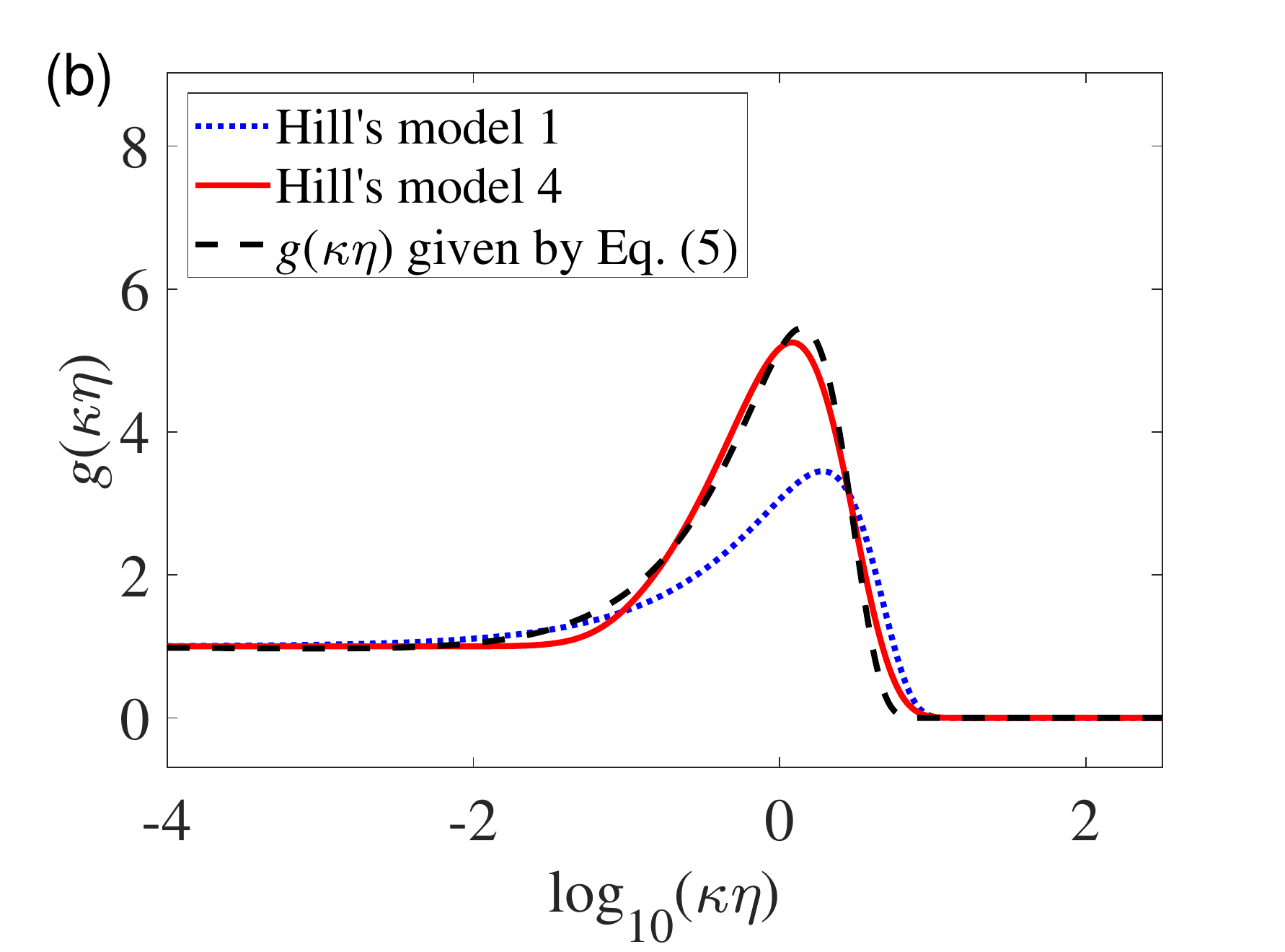}\\
	\includegraphics[width=0.48\textwidth]{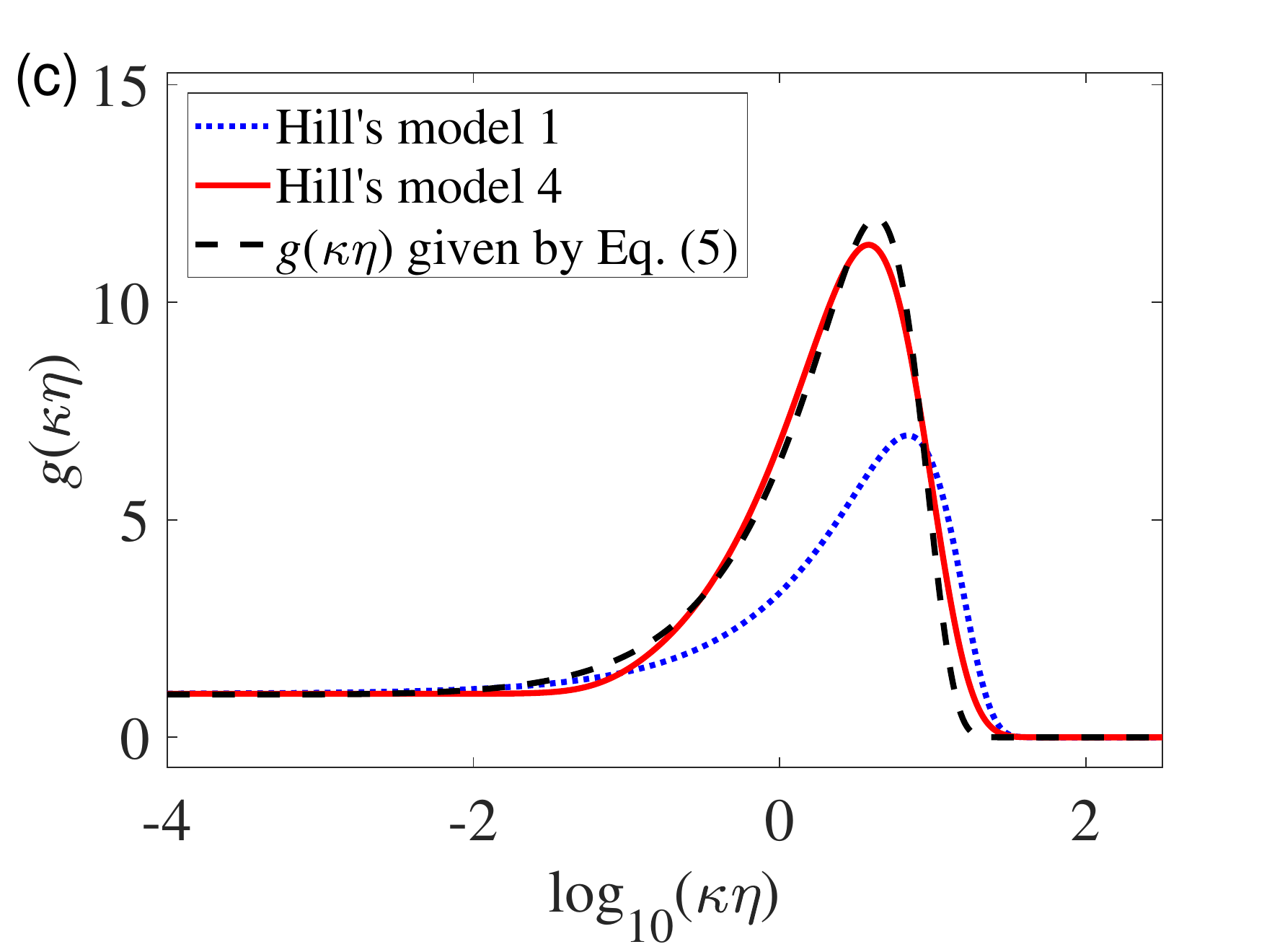}
	\includegraphics[width=0.48\textwidth]{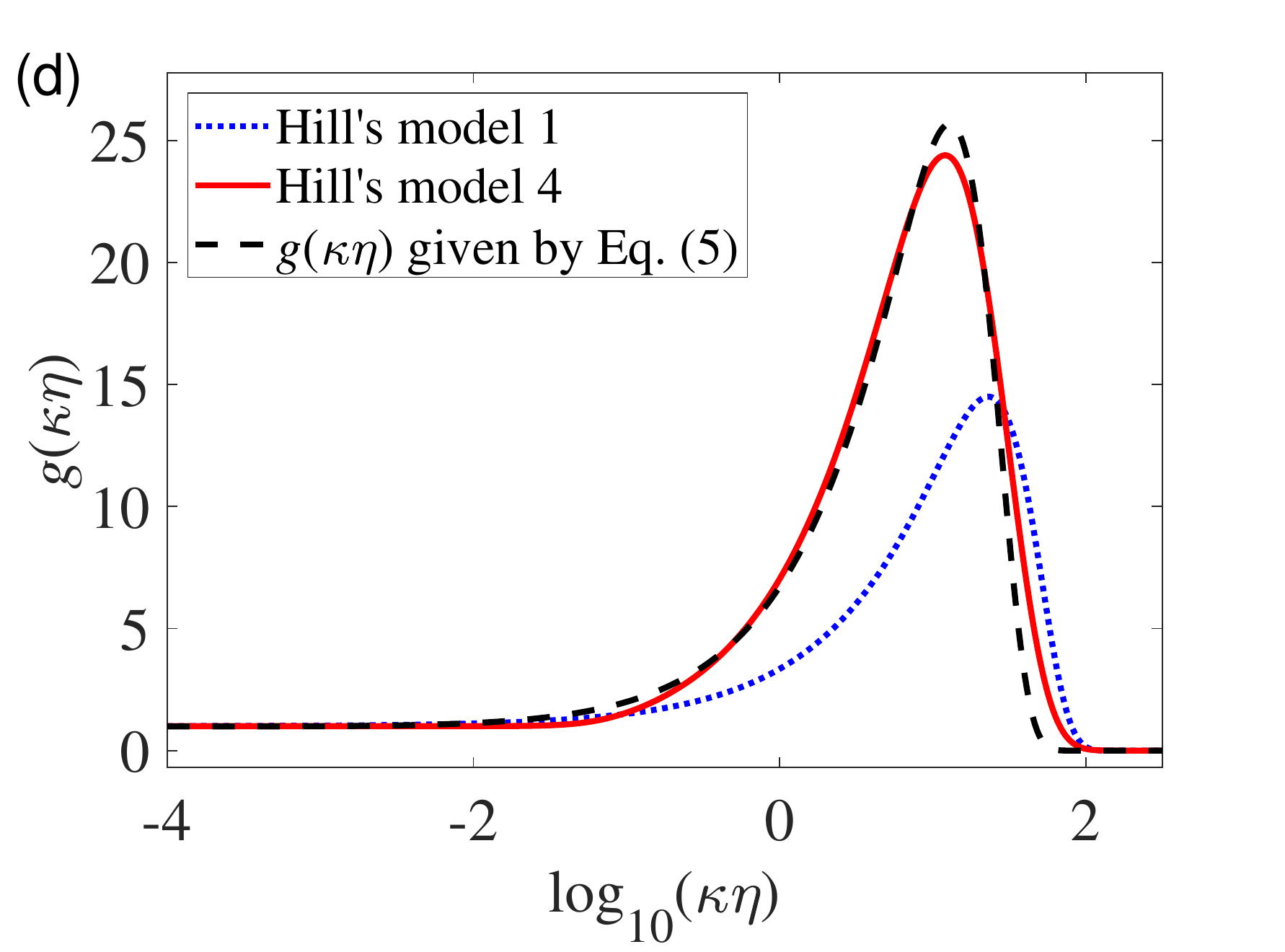}
	\caption{Universal dimensionless function $g(\kappa\eta)$ at different Prandtl numbers: (a) $Pr=3$, (b) $Pr=30$, (c) $Pr=300$, (d) $Pr=3000$.}
	\label{fig2}
\end{figure*}

To examine the accuracy of the fitted solution for $g(\kappa\eta)$, we compare it with analytic formula H1 (see Appendix I) and H4 (calculated numerically) for several values of the Prandtl number. 
Figure \ref{fig1} includes comparison among the H1 model, H4 model, our analytic fit and model in \cite{Yi:18} at $Pr=7$ and $Pr=700$. At these values our fit is in a very good agreement with the H4 model and the model in \cite{Yi:18}. Figure \ref{fig2}(a)-\ref{fig2}(d) illustrate the ability of our analytic fit to match H4 model within the wide range [3,3000] of the Prandtl numbers. Figure \ref{fig2}(a) produced for $Pr=3$ implies that the fitted function $g(\kappa\eta)$ calculated from Eq. (5) and that obtained from H1 model agree with that based on H4 model reasonably well while Eq. (\ref{eq5}) provides with a better bump-shape reconstruction. Further, Fig. \ref{fig2}(b)-\ref{fig2}(d), obtained for $Pr=30, 300$ and $3000$, respectively, illustrate that $g(\kappa\eta)$ in Eq.  (5) is very consistent with that based on H4, but it is not the case for $g(\kappa\eta)$ based on H1, the latter leading to drastic underestimation of the bump's strength. Hence, if one uses H4 as a benchmark, the proposed $g(\kappa \eta)$ in Eq.  (5) and, hence, the ${\Phi}(\kappa )$ in Eq. (\ref{eq_a1}) show advantages in describing the spectral bump for all $Pr \in [3, 3000]$, and especially so for $Pr \geqslant 30$. 

We note that $g(\kappa\eta)$ in Eq. (\ref{eq5}) has the same functional form as those in the Andrews' \cite{Andrews_1992} and in the Grayshan's \cite{Grayshan2007} power spectra being the product of a polynomial and a Gaussian function.  On the other hand, it also relies on $\kappa$ and $Pr$ independently, making the bump height ${g_{\max }}$ increase with growing $Pr$ values, which is of importance in achieving the precision in the spectrum model valid for the wide range of $Pr$.

Thus, most existing H4-based models such as Yi's model \cite{Yi:18} and Andrews' model \cite{Andrews_1992}, use fixed Prandtl number and/or fixed Schmidt number. These models are very accurate. For analysis of power spectra with $Pr_{\rm{T}} = 7$ and Schmidt number $Pr_{\rm{S}} = 700$, model in Ref. \cite{Yi:18} can be used. However, as we have clarified, the diversity of oceanic environment leads to drastic deviation from values $Pr_{\rm{T}} = 7$ and $Pr_{\rm{S}} = 700$. In such cases model in Eq. (\ref{eq_a1}) covering range $[3,3000]$ becomes very convenient.

\subsection{Dissipation constraint analysis}
\label{Sec_1b}
Although a good numerical fit for the power spectrum was obtained in the previous section for a wide range of $Pr$, it appears of importance to estimate whether the new model agrees with the basic laws of fluid mechanics. In this section we will verify to which degree the dissipation constraint \cite{Frehlich_1992,Muschinski_2015TemVa} being the consequence of the scalar transport equation is satisfied by Eq. (\ref{eq5}). The dissipation constraint can be expressed via integral \cite{Frehlich_1992}:
\begin{equation}
{\rm X} = \frac{{2\beta }}{{Pr }}\int_0^\infty  {g(x)x^{\frac{1}{3}}{\rm{d}}x} .
\label{eq6}
\end{equation}
The closer {\rm X} is to 1, the more $g(x)$ agrees with the dissipation constraint.

\begin{figure*}[b!]
	\centering
	\includegraphics[width=0.6\textwidth]{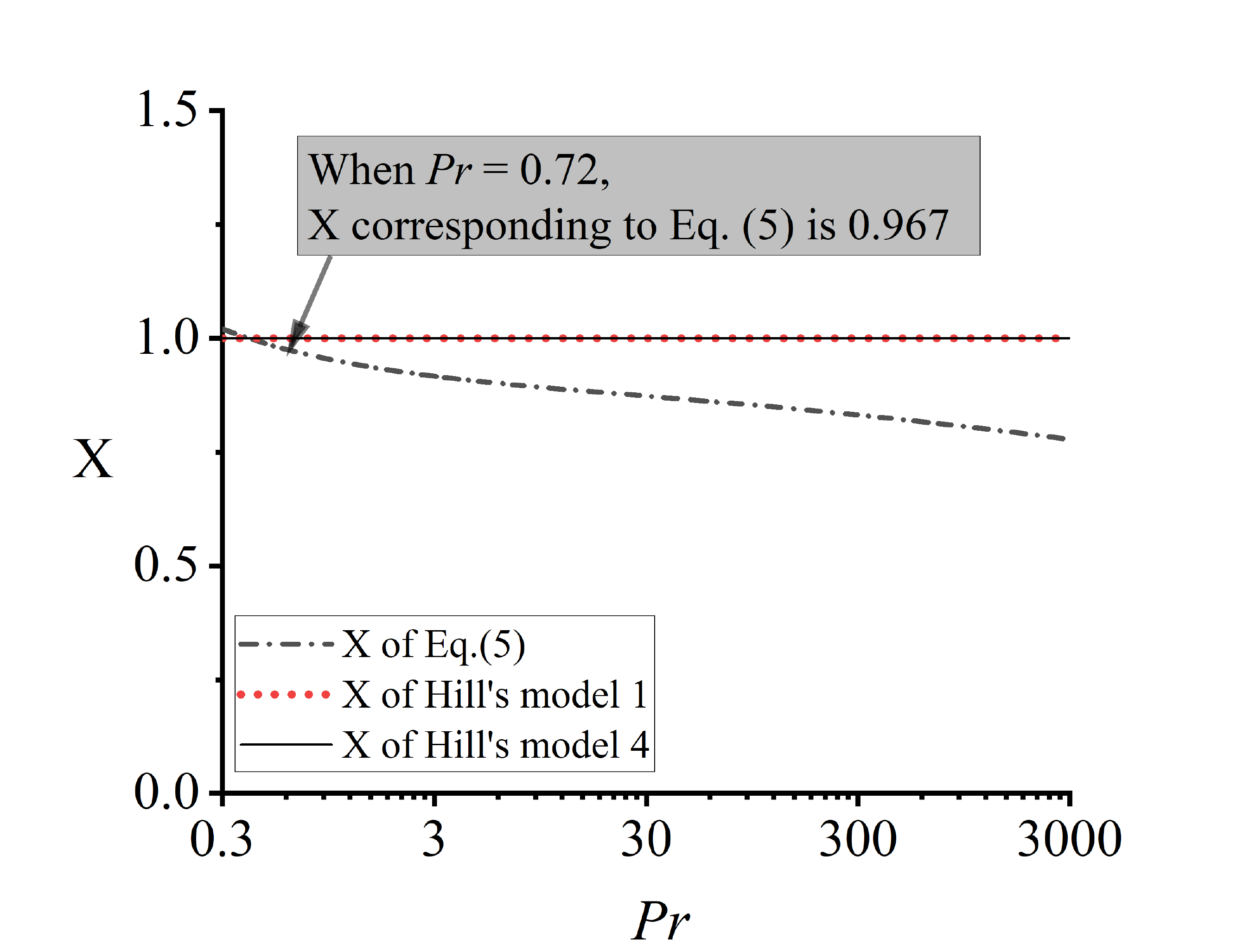}
	\caption{The values of X vary with Prandtl number in H1 model, H4 model and Eq. (5).}
	\label{fig_X}
\end{figure*}

The values of ${\rm X}$ calculated on substituting Eq. (5) into Eq. (7) are shown in Table \ref{tab1} and Fig. \ref{fig_X} for different values of the  Prandtl number. The table entries imply that Eq.  (5) underpredicts {\rm X} by $8.3\% $ and $22.2\% $ when $Pr = 3$ and $3000$, respectively.
This discrepancy is intensified with the increasing Prandtl number which, in its turn, corresponds to the increasing bump height. This indicates that a more precise fitting of the spectral bump occurs at the expense of deviation from the dissipation constraint.

\begin{table}[h]
	\centering
	\caption{\bf Values of ${\rm X}$ calculated from Eq. (\ref{eq5}) for various Prandtl numbers.}
	\setlength{\tabcolsep}{8pt}
	\begin{tabular}{ccccc}
		\hline
		${Pr}$ & $3$ & $30$ & $300$ & $3000$ \\
		\hline
		${\kern 5pt}{\rm X}{\kern 5pt}$ & $0.917$ & $0.873$ & $0.831$ & $0.778$ \\
		\hline
	\end{tabular}
	\label{tab1}
\end{table}

For comparison,
${\rm X}$ corresponding to other models were calculated in \cite{Muschinski_2015TemVa}, and are shown in Table \ref{tab2}. 
These models fit H4 at $Pr = 0.72$. 
Apart from the Frehlich model, the values of X for the rest of the models do exhibit certain deviation from value 1. On comparing entries of Tables 1 and 2 it appears that the model in Eq. (\ref{eq5}) meets the dissipation constraint in a satisfactory manner.
\begin{table}[h]
	\centering
	\caption{\bf The values of ${\rm X}$ corresponding to reported models}
	\setlength{\tabcolsep}{4.5 pt}
	\begin{tabular}{ccccc}
		\hline
		Models & \emph{Frehlich}\cite{Frehlich_1992} & \emph{Churnside}\cite{Churnside1990A} & \emph{Andrews}\cite{Andrews_1992} & \emph{Grayshan}\cite{Grayshan2007} \\
		\hline
		${\rm X}$ for $Pr = 0.72$ & $1.00$ & $1.12$ & $0.95$ & $1.37$ \\
		\hline
	\end{tabular}
	\label{tab2}
\end{table}

\textit{The power spectrum model given by Eq. (\ref{eq_a1}) is the first of the two main results of our paper.} It can be used for any turbulent medium in which a single scalar is advected by the velocity field and is applicable for the wide range of the Prandtl numbers, $Pr\in [3,3000]$. Moreover, it is confirmed that  model (\ref{eq_a1}) is in the reasonable agreement with the dissipation constraint. 

\section{The two-scalar oceanic refractive-index spectrum and optical scintillation}
\label{Sec_2}
\subsection{Practical example: oceanic refractive-index spectrum }
\label{Sec_2a}

In this section we will demonstrate how to extend the power spectrum model given by Eq. (\ref{eq_a1}) from one to two advected scalars with different Prandtl numbers and Schmidt numbers. The best illustration of such a medium is the oceanic turbulence, hence we will adopt the corresponding notations to maintain practicality. 

On following the Nikishov's linearized polynomial approach \cite{Nikishov2000Spectrum} we assume that the fluctuating portion of the refractive index $n$ is given by linear combination 
\begin{align}
n' =  - AT' + BS'
\label{eq7}
\end{align}
of the temperature fluctuation $T'=T-\langle T \rangle$ and the salinity concentration fluctuation $S'=S-\langle S \rangle$, $T$ and $S$ representing the instantaneous temperature and salinity values and the angular brackets standing for the statistical average. In Eq. (\ref{eq7}) $A$ is the thermal expansion coefficient and $B$ is the saline contraction coefficient \cite{Nikishov2000Spectrum}. 

Therefore, the power spectrum of the refractive index fluctuations can be expressed as the following linear combination of the temperature spectrum ${\Phi _{\rm{T}}}(\kappa )$, the 
salinity spectrum ${\Phi _{\rm{S}}}(\kappa )$, and the co-spectrum ${\Phi _{{\rm{TS}}}}(\kappa )$ \cite{Nikishov2000Spectrum}: 
\begin{align}
{\Phi _{\rm{n}}}(\kappa ) = {A^2}{\Phi _{\rm{T}}}(\kappa ) + {B^2}{\Phi _{\rm{S}}}(\kappa ) - 2AB{\Phi _{{\rm{TS}}}}(\kappa ).
\label{eq8}
\end{align}
However, unlike in \cite{Nikishov2000Spectrum} where each of these three spectra was based on the H1 model, we will use the fit for the H4 model obtained above [see Eq. (\ref{eq_a1})], i.e.,   
\begin{align}
	\nonumber {\Phi _i}(\kappa ) = &\left[ {1 + {\rm{21}}{\rm{.61}}{{(\kappa \eta )}^{{\rm{0}}{\rm{.61}}}}{c_i}^{{\rm{0}}{\rm{.02}}} - {\rm{18}}{\rm{.18}}{{(\kappa \eta )}^{0.55}}{c_i}^{{\rm{0}}{\rm{.04}}}} \right]\\
	& \times\frac{1}{{4\pi}}\beta {\varepsilon ^{ - \frac{1}{3}}}{\kappa ^{ - \frac{11}{3}}}{\chi _i} \exp \left[ { - 174.90{{(\kappa \eta )}^2}{c_i}^{{\rm{0}}{\rm{.96}}}} \right], \quad i \in \{{\rm{T}},{\rm{S}},{\rm{TS}}\} , 
	\label{eq9}
\end{align}
where ${c_i} = {0.072^{4/3}\beta {Pr }_i^{-1}}$, ${Pr }_{\rm{T}}$ and ${Pr }_{\rm{S}}$ are the temperature Prandtl number and salinity Schmidt number, respectively,
${Pr _{\rm{TS}}} = 2{Pr _{\rm{T}}}{Pr _{\rm{S}}}({Pr _{\rm{T}}}+{Pr _{\rm{S}}})^{-1}$ is the coupled Prandtl-Schmidt number \cite{JinrenYao_17,Yi:18}, and
${\chi _i}$ are the ensemble-averaged variance dissipation rates which 
are related by expressions \cite{Elamassie:17}
\begin{align}
{\chi _{\rm{S}}} = \frac{{{A^2}}}{{{\omega ^2}{B^2}}}{\chi _{\rm{T}}}{d_r}, \quad {\chi _{{\rm{TS}}}} = \frac{A}{{2\omega B}}{\chi _{\rm{T}}}\left( {1 + {d_r}} \right),
\label{eq10}
\end{align}
${d_r}$ being the eddy diffusivity ratio and $\omega $ being the relative strength of temperature- salinity fluctuations.
In most practical cases, 
${d_r}$ and $\omega $ are related as \cite{Elamassie:17}
\begin{align}
	{d_r} \approx \left\{ {
	\begin{array}{ll}
	{\left| \omega  \right|} + {\left| \omega  \right|^{0.5}}{\left( {\left| \omega  \right| - 1} \right)^{0.5},} & {\left| \omega  \right| \ge 1}\\
	{1.85\left| \omega  \right| - 0.85,} & {0.5 \le \left| \omega  \right| < 1}\\
	{0.15\left| \omega  \right|,} & {\left| \omega  \right| < 0.5}
	\end{array}}
\right..
\label{eq07_3}
\end{align}

\textit{The power spectrum model given by Eqs. (\ref{eq8}) and (\ref{eq9}) is the second of the two main results of our paper.} It gives the analytic description of optical turbulence with two advected quantities while each of them may have Prandtl or Schmidt numbers in interval  $[3,3000]$. In its present form the new model  is specifically applied for the oceanic waters in which the turbulent velocity field advects temperature and salinity concentration. However, it can be readily adapted for a variety of other turbulent media based on two scalars, for instance a fresh water contaminated by a chemical substance or any exotic turbulent fluid. Furthermore, a straightforward linearization of three or more advected quantities can be worked out in a similar manner.  

\subsection{The scintillation index of a spherical wave}
\label{Sec_2b}

We will now illustrate how the analytical fit for the oceanic power spectrum given by Eqs. (\ref{eq8}) and (\ref{eq9}) compares with the Nikishovs' spectrum (based on analytical model H1) and the H4 model (handled via numerical calculations) as it manifests itself in the evolution of the scintillation index of light propagating though an extended turbulent channel. The scintillation index of an optical wave at distance $L$ from the source plane is generally defined by expression \cite{Phillips2005Laser}
\begin{equation}
\sigma^2_I(L)=\frac{\langle I^2(L)\rangle}{\langle I(L)\rangle^2}-1,
\end{equation}
where $I$ is the instantaneous intensity of the wave, and the angular brackets stand for the ensemble average. 

We will restrict ourselves to the scintillation index of a spherical wave being one of the most important parameters of light-turbulence interaction. In addition to the Rytov variance, i.e., the scintillation index of the plane wave, it can serve as an indicator of the global turbulence regime (weak/focusing/strong) \cite{Phillips2005Laser}. This quantity was extensively studied for the atmospheric propagation in dependence from a number of turbulence parameters \cite{Phillips2005Laser}. Based on the Nikishovs' spectrum the scintillation index of the spherical wave was also previously evaluated on propagation in the oceanic water (c.f. \cite{FarwellOKSpectrum}).

Let us now analytically derive the expression for the scintillation index of the spherical wave in the water channel described by spectrum in Eqs. (\ref{eq8}) and (\ref{eq9}). According to the Rytov perturbation theory it takes form
\begin{align}
	\nonumber \sigma^2_I(L)= &{\kern 3pt} 4\pi {\mathop{\rm Re}\nolimits} \biggl\{{\int_0^L {{\rm{d}}\varsigma \int_0^\infty  {\kappa {\rm{d}}\kappa \int_0^{2\pi } {{\rm{d}}\theta \bigg[ {{{\left| {E\left( {\varsigma ,\kappa ,L} \right)} \right|}^2}}} } } }\\
	& \left. { { + E\left( {\varsigma ,\kappa ,L} \right) \times E\left( {\varsigma ,\kappa ,L} \right)}\bigg] \frac{{{\Phi _{\rm{n}}}(\kappa )}}{{{n_0}^2}}} \right\},
	\label{eq12}
\end{align}
with
\begin{align}
E\left( {\varsigma ,\kappa ,L} \right) = {\rm{i}}k\exp \left[ { - \frac{{0.5{\rm{i}}\varsigma (L - \varsigma )}}{{kL}}{\kappa ^2}} \right],
\label{eq13}
\end{align}
where $ \rm{i} = \sqrt{-1}$ (not to be confused with subscript $i$ also used in the paper),
$k = {{2\pi {n_0}} \mathord{\left/
		{\vphantom {{2\pi {n_0}} {{\lambda _0}}}} \right.
		\kern-\nulldelimiterspace} {{\lambda _0}}}$, ${\lambda _0}$ is the wavelength in vacuum, 
${n_0}$ (ca. $1.33$) is the averaged refractive-index of the ocean water.

On substituting from Eq. (\ref{eq8}) into Eqs. (\ref{eq12}) and (15) we arrive at the scintillation index in of the form
\begin{align}
	\nonumber \sigma^2_I(L) = &\left( {{A^2}{\chi _{\rm{T}}}{M_{\rm{T}}} + {B^2}{\chi _{\rm{S}}}{M_{\rm{S}}} - 2AB{\chi _{{\rm{TS}}}}{M_{{\rm{TS}}}}} \right)\\
	&\times\pi L\beta {\varepsilon ^{ - \frac{1}{3}}}{\eta ^{\frac{5}{3}}}{\left( {\frac{k}{{{n_0}}}} \right)^2},
	\label{eq14}
\end{align}
with
\begin{align}
	\nonumber{M_i} = &\sum\limits_{\ell = 1}^3 {{P_{\ell,1}}{c_i}^{{P_{\ell,2}}}{{\left( {174.90{c_i}^{0.96}} \right)}^{(\frac{5}{6} - \frac{{{P_{\ell,3}}}}{2})}}\Gamma \left( {\frac{{{P_{\ell,3}}}}{2}} - \frac{5}{6}\right)} \\
	&\nonumber\times \left[ {1 - {}_3{F_2}\left( {1,\frac{{{P_{\ell,3}}}}{4} - \frac{5}{{12}},\frac{{{P_{\ell,3}}}}{4} - \frac{1}{{12}};\frac{3}{4},\frac{5}{4};} \right.} \right.\\
	&\left. {\left. {\frac{{-{L^2}}}{{16 \times {{\left( {174.90{c_i}^{0.96}} \right)}^2}{k^2}{\eta ^4}}}} \right)} \right], \quad \{i=\rm{T}, \rm{S}, \rm{TS}\}
	\label{eq15}
\end{align}
and 
\begin{align}
{\textbf{P}} = \{ P_{ij}\}= \left( {\begin{array}{*{20}{c}}
	1&0&0\\
	{{\rm{21}}{\rm{.61}}}&{0.02}&{0.61}\\
	{ - {\rm{18}}{\rm{.18}}}&{0.04}&{0.55}
	\end{array}} \right).
\label{eq16}
\end{align}
Here $\Gamma$ is the Gamma function and ${}_3{F_2}$ is the generalized Hypergeometric function \cite{MathHB}. Equations (\ref{eq14})-(\ref{eq16}) combined with the Eqs. (\ref{eq10}) and (\ref{eq07_3}) imply that  
$\sigma^2_I$ is proportional to $\chi _{\rm{T}}$ and ${\varepsilon ^{ - 1/3}}$ but show a somewhat complex dependence on $\omega$, ${Pr _i}$ and $\eta$.

\begin{figure*}[!b]
	\centering
	\includegraphics[width=0.48\textwidth]{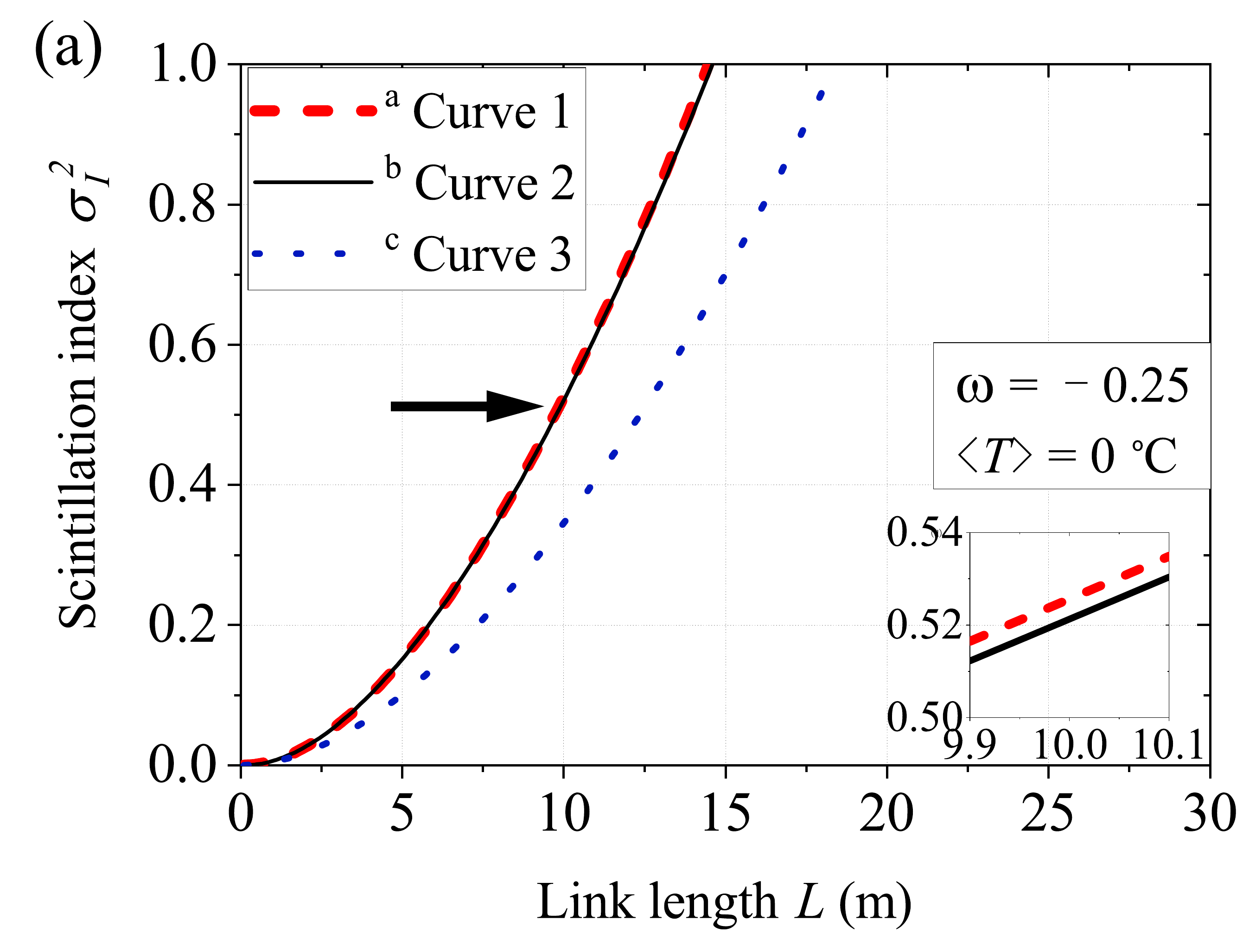}
	\includegraphics[width=0.48\textwidth]{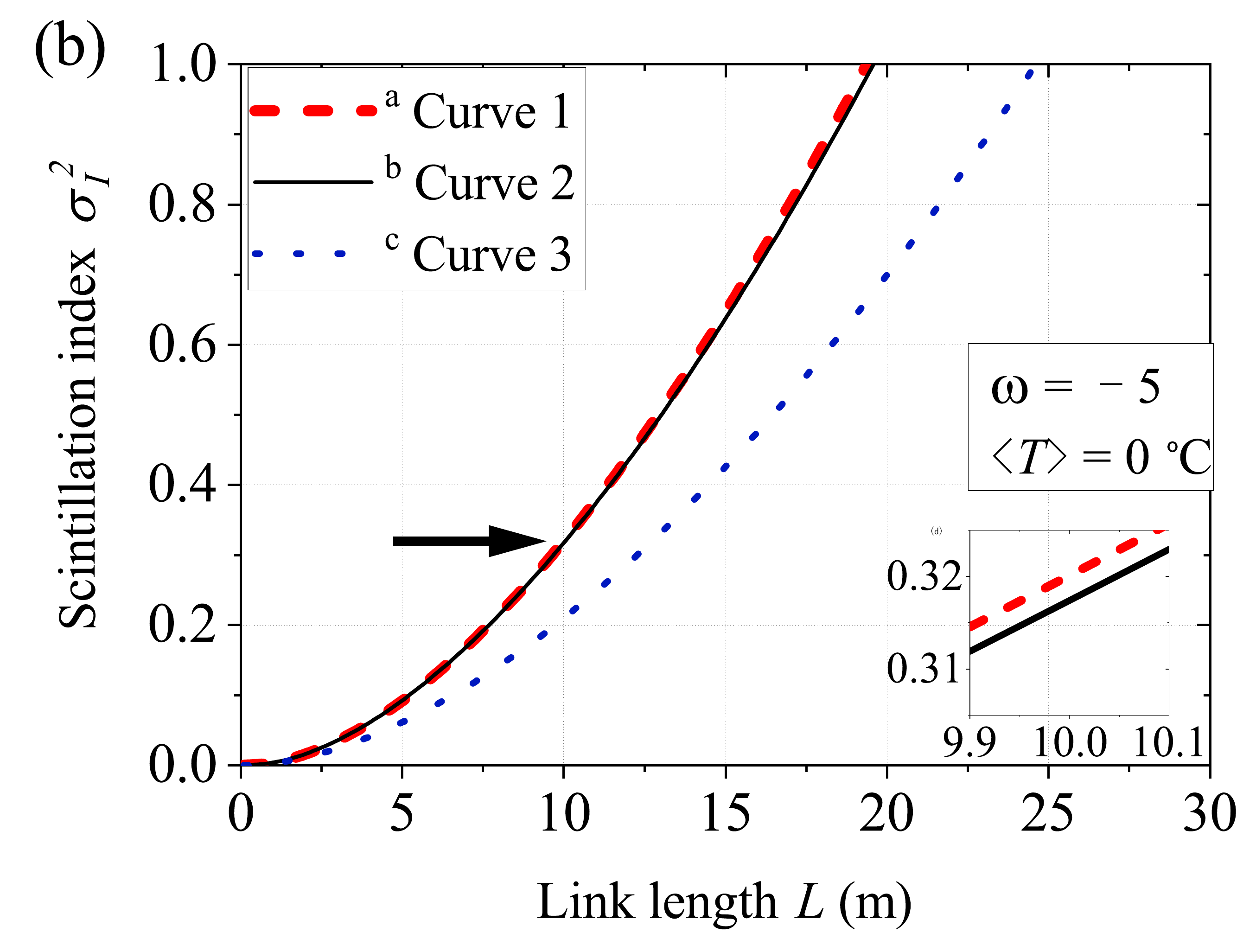}
	\includegraphics[width=0.48\textwidth]{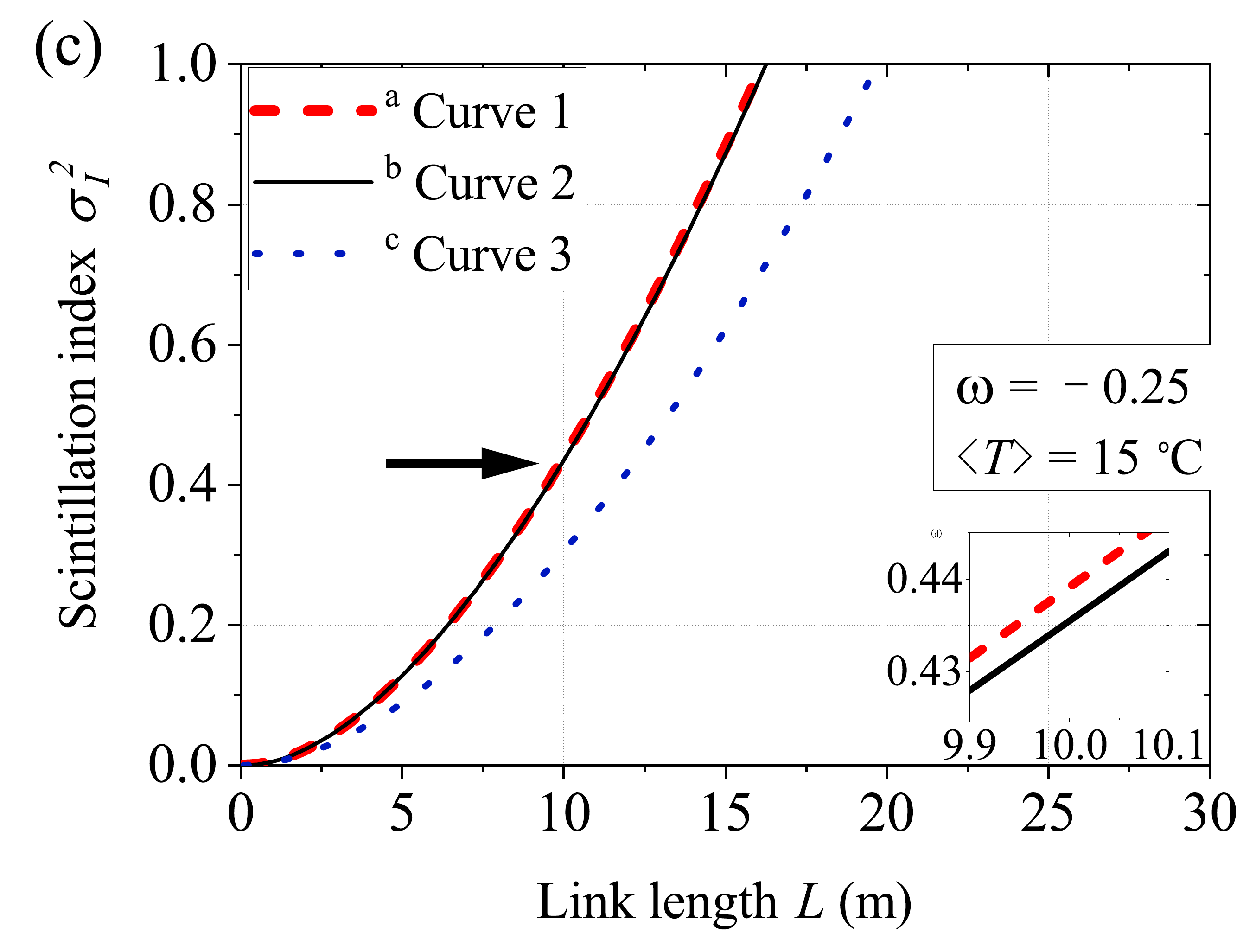}
	\includegraphics[width=0.48\textwidth]{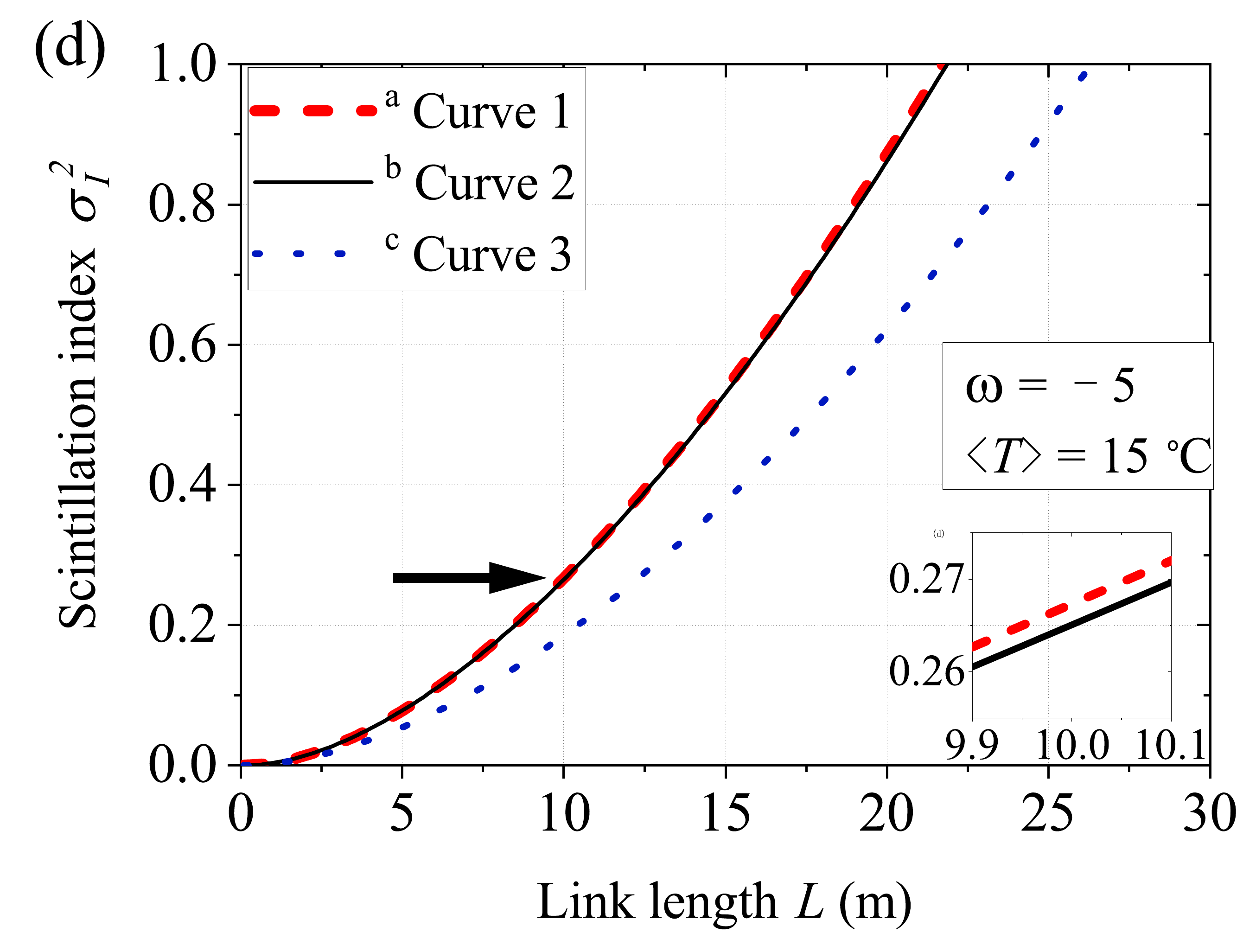}
	\includegraphics[width=0.48\textwidth]{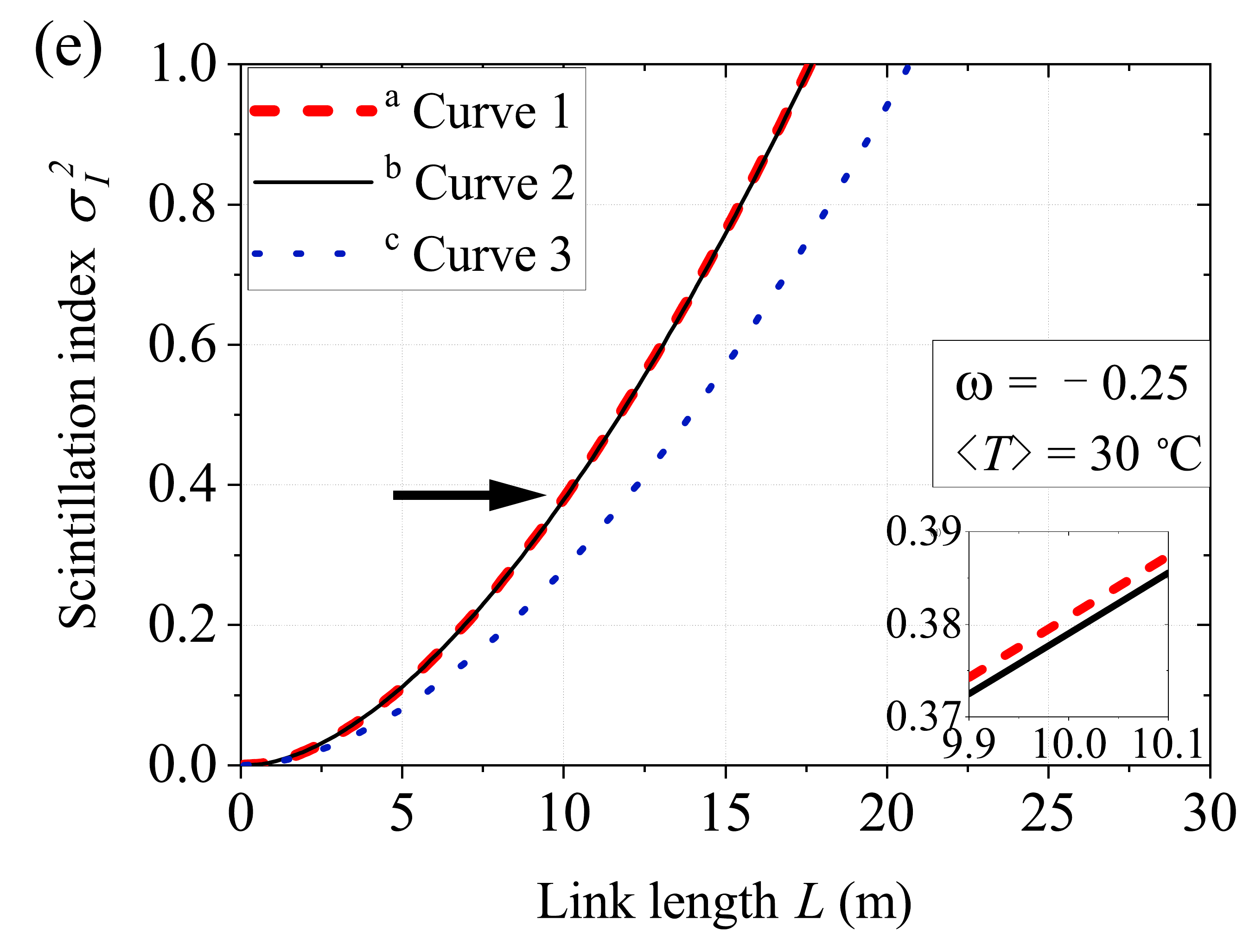}
	\includegraphics[width=0.48\textwidth]{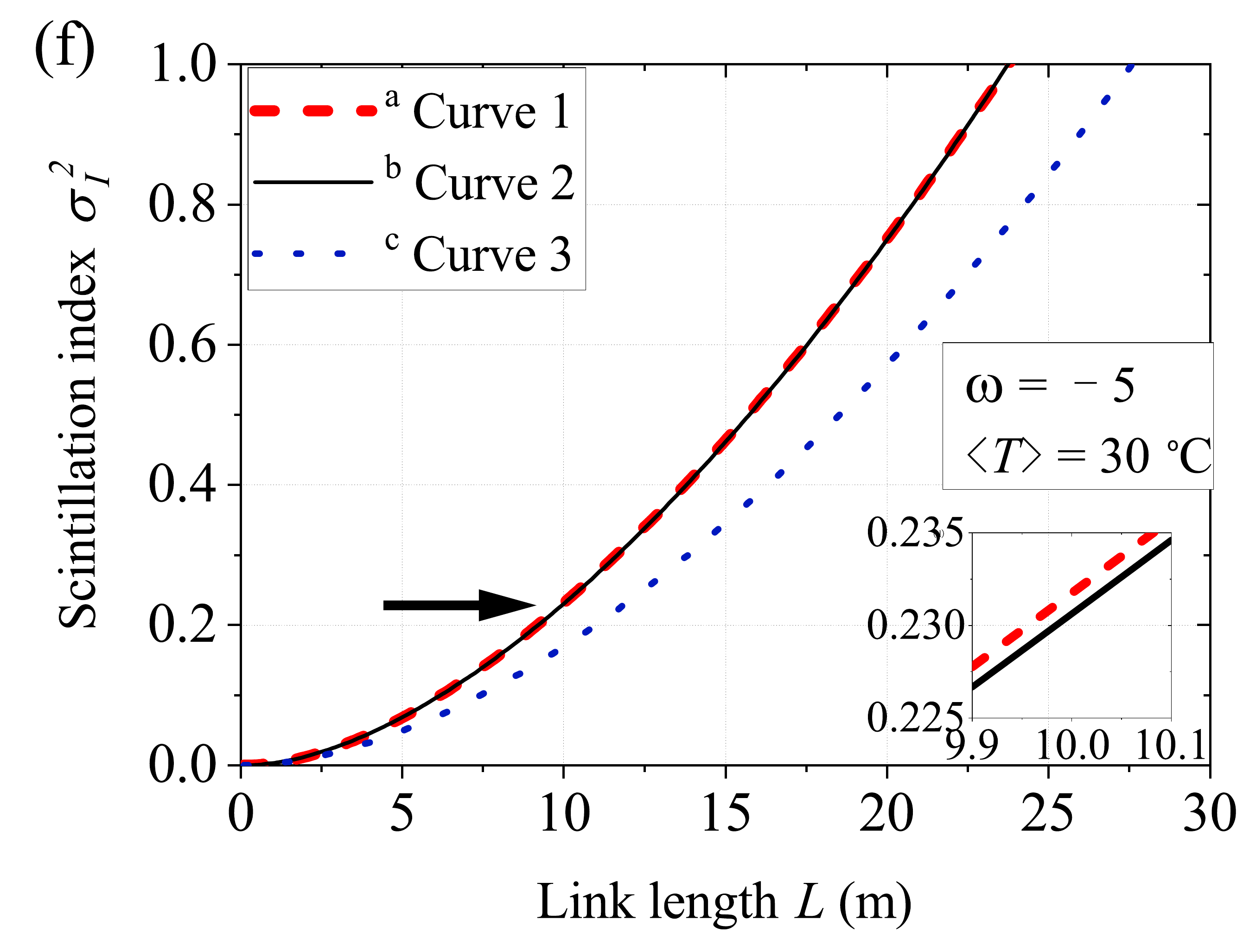}
	\caption{The scintillation index of a spherical wave corresponding to different power spectrum models, for several different values of $\omega$ and $\left\langle T \right\rangle$: (a) $\omega = -0.25$, $\left\langle T \right\rangle = 0^{\circ}\rm C$, (b) $\omega = -5$, $\left\langle T \right\rangle = 0^{\circ}\rm C$, (c) $\omega = -0.25$, $\left\langle T \right\rangle = 15^{\circ}\rm C$, (d) $\omega = -5$, $\left\langle T \right\rangle = 15^{\circ}\rm C$, (e) $\omega = -0.25$, $\left\langle T \right\rangle = 30^{\circ}\rm C$, (f) $\omega = -5$, $\left\langle T \right\rangle = 30^{\circ}\rm C$.
		$^a$ Curve 1 (red, dashed curve) is calculated from Hill's model 4 numerically. 
		$^b$ Curve 2 (black, solid curve) is based on Eq. (\ref{eq14}). 
		$^c$ Curve 3 (blue, dotted curve) is calculated from Nikishovs' spectrum numerically.
	}
	\label{fig4}
\end{figure*}

In order to compare the values of $\sigma^2_I(L) $ corresponding to different oceanic power spectrum models, 
we set ${n_0} = 1.33$, ${\lambda _0} = 532{\rm{nm}}$, 
$\beta  = 0.72$,
$A = 2.{\rm{5}}6 \times {10^{ - 4}}{\deg ^{ - 1}}{\rm{l}}$,  
$B = {\rm{7}}{\rm{.44}} \times {10^{ - 4}}{{\rm{g}}^{ - 1}}{\rm{l}}$,  
${\chi _{\rm{T}}} = 1 \times {10^{ - 5}}{{\rm{K}}^{\rm{2}}}{{\rm{s}}^{{\rm{ - 1}}}}$, 
$\varepsilon  = 1 \times {10^{ - 2}}{{\rm{m}}^{\rm{2}}}{{\rm{s}}^{{\rm{ - 3}}}}$,  
but vary $\omega$ and average temperature $\left\langle T \right\rangle$. Varying $\left\langle T \right\rangle$ leads to the changes of ${Pr _i}$ and $\eta$ (more details in the Appendix II). In figure \ref{fig4} the scintillation index $\sigma^2_I(L) $ calculated from Eq.  (\ref{eq14}) is plotted for different values of $\omega$ and $\left\langle T \right\rangle$ (solid black curve or curve 2) and is compared with that calculated from the H4 model (dashed red curve or curve 1) and also with that given by the Nikishovs' H1-based model (dotted blue curve or curve 3). Figure \ref{fig4} suggests that if H4 is considered as the most precise model, a standard, then the scintillation index given in Eq. (\ref{eq14}) provides a very accurate approximation to that based on H4, which is not the case for that based on the Nikishovs' model. Such an advantage stems from the fact that the information about the bumps occurring in all three spectra, $\Phi_{\rm{T}}$, $\Phi_{\rm{S}}$ and $\Phi_{\rm{TS}}$, at high spatial wave numbers has been taken into the account by Eq. (\ref{eq14}) at its best.

As Fig. \ref{fig4} indicates, the average temperature is not in the direct relation with the scintillation index. Indeed, the scintillation index is directly affected by the variance of the fluctuating temperature rather than by its average value.

\section{Summary}
\label{Sec_3}
The power spectrum of the refractive index fluctuations is an effective tool for optical characterization of the statistics of turbulent medium (up to the second order) in its homogeneous regime. The widely accepted model H4 for the power spectrum of a single scalar advected by the turbulent fluid is based on a nonlinear differential equation and can always be used for numerical calculation of the spectrum. However, for light propagation problems it is desirable to have an accurate yet tractable analytical model for the power spectrum. For the turbulent media with low Prandtl numbers, as is the case for the atmospheric turbulence, such analytical models have been developed long time ago by fitting an analytic curve to the H4 profile. However, for turbulent media in which the scalar fields are advected at high and possibly variable Prandtl numbers, the accurate analytical fits have not yet been introduced. 

In this paper we have introduced a very simple yet accurate analytical model for the power spectrum of turbulence with either one or two advected scalars. We have illustrated our procedure for the best example of such a medium - oceanic turbulence - and have obtained a very good agreement between such power spectrum and that based on the H4 numerical curve. We have also shown that the Nikishovs' model widely used to characterize oceanic optical turbulence is much less accurate. As a consequence, our model's close fit to the H4-based spectrum result in the close agreement for the statistics of the propagating light. In particular, we have illustrated that the scintillation index of a spherical wave as calculated from our power spectrum model is practically indistinguishable from that based on the H4 model. At the same rate, the Nikishovs' spectrum based scintillation index is largely underestimated.  

The key difference between our new power spectrum and the previously introduced ones stems from the fact that in our model the Prandtl/Schmidt numbers of the advected scalars are given as parameters that may vary independently in the interval $Pr\in [3,3000]$. Such a large range produces a slight deviation of the new spectrum from the H4 model for all the $Pr$ values which can be substantially reduced if a smaller range is set. However, this very large range of the $Pr$ values leads to our model's main advantage over all other models developed so far: the intrinsic flexibility in accounting for various thermodynamic states. Since, in general, the Prandtl/Schmidt numbers are very sensitive to a number of factors, such as the average temperature, pressure, concentration, etc., it is of utmost importance to have a power spectrum being flexible enough to account for these variations. Moreover, this very feature becomes of fundamental convenience when one may want to apply our model for homogeneous turbulence developed in a fluid other than the oceanic water. Hence we envision that our development may find uses in ecology, meteorology, food processing, medicine and some other technologies that use light propagation in a variety of turbulent media.

\section*{Appendix I: Nikishovs' H1-based model}
\label{Ap_1}
Here we summarize the H1 model and the Nikishov's model based on it (see  \cite{hill_1978,Nikishov2000Spectrum,KOROTKOVA_Lightocean,Elamassie:17} for more details).

H1 describes the three-dimensional scalar power spectrum of a single advected quantity as
\begin{align}
\Phi (\kappa ) = \frac{1}{{4\pi }}\beta \chi {\varepsilon ^{ - \frac{1}{3}}}{\kappa ^{ - \frac{11}{3}}}g(\kappa \eta),
\label{eq_ap1}
\end{align}
where the universal dimensionless function $g(\kappa \eta)$ is set to 
\begin{align}
g(k\eta ) = \left[ {1 + Q{{(k\eta )}^{\frac{2}{3}}}} \right]\exp \left[ { - \beta \frac{{\frac{3}{2}{Q^2}{{(k\eta )}^{\frac{4}{3}}} + {Q^3}{{(k\eta )}^2}}}{{{Q^2}Pr}}} \right].
\label{eq_ap2}
\end{align}
Nikishovs' power spectrum model for the refractive index fluctuations is the linear combination of the  temperature spectrum, the salinity spectrum and their co-spectrum:
\begin{align}
{\Phi _{\rm{n}}}(\kappa ) = {A^2}{\Phi _{\rm{T}}}(\kappa ) + {B^2}{\Phi _{\rm{S}}}(\kappa ) - 2AB{\Phi _{{\rm{TS}}}}(\kappa ),
\label{eq_ap3}
\end{align}
where each term is given by Eqs.  (\ref{eq_ap1}) and  (\ref{eq_ap2}), i.e., 
\begin{align}
\nonumber {\Phi _i}(\kappa ) = &{\rm{ }}\frac{1}{{4\pi }}\beta {\varepsilon ^{ - \frac{1}{3}}}{\chi _i}{\kappa ^{ - \frac{{11}}{3}}}\left[ {1 + Q{{\left( {\kappa \eta } \right)}^{\frac{2}{3}}}} \right]\\
& \times \exp \left[ { - \beta \frac{{\frac{3}{2}{Q^2}{{\left( {\kappa \eta } \right)}^{4/3}} + {Q^3}{{\left( {\kappa \eta } \right)}^2}}}{{{Q^2}{{Pr }_i}}}} \right], \quad \{i=\rm{T}, \rm{S}, \rm{TS}\}.
\label{eq_ap4}
\end{align}

\section*{Appendix II: Oceanic turbulence parameters depending on average temperature}
\label{Ap_2}
Table \ref{tab3} gives the values of several average temperature-based parameters in oceanic refractive-index spectrum when salinity $\left\langle S \right\rangle =34.9\rm{ppt}$.
The values of kinematic viscosity $\upsilon$ and temperature Prandtl number ${Pr _{\rm{T}}}$ are calculated with the codes available in http://web.mit.edu/seawater/ \cite{MIT:r1,MIT:r2}. 
The values of salinity diffusivity $\alpha _{\rm{S}}$ for $\left\langle T \right\rangle =15^{\circ}\rm C$ and $30^{\circ}\rm C$ are obtained from the diffusivity of $\rm{Cl}^-$  \cite{Poisson1983};
the $\alpha _{\rm{S}}$ for $\left\langle T \right\rangle =0^{\circ}\rm C$ is calculated from that for $15^{\circ}\rm C$, $20^{\circ}\rm C$,	$25^{\circ}\rm C$, $27^{\circ}\rm C$ and $30^{\circ}\rm C$ by the Stokes--Einstein law \cite{Poisson1983}. 
We calculate the Schmidt number by ${Pr _{\rm{S}}} = {\nu  \mathord{\left/
		{\vphantom {\nu  {{\alpha _{\rm{S}}}}}} \right.
		\kern-\nulldelimiterspace} {{\alpha _{\rm{S}}}}}$ and the Kolmogorov microscale by $\eta  = {\nu ^{3/4}}{\varepsilon ^{ - 1/4}}$.
\begin{table}
	\centering
	\caption{\bf Values of viscosity, diffusivity and Prandtl/Schmidt  number in different values of temperature when salinity is $34.9\rm{ppt}$}
	\newcommand{\tabincell}[2]{\begin{tabular}{@{}#1@{}}#2\end{tabular}}
	\begin{tabular}{cccccccccc}
		\hline
		 \tabincell{c}{Average salinity  $\left\langle S \right\rangle$  ($\rm{ppt}$)}  & & & \multicolumn{5}{c}{34.9} & & \\
		\hline
		
		\tabincell{c}{Average temperature  $\left\langle T \right\rangle$  ($^{\circ}\rm C$)}  & & & 0 & & 15 & & 30 & & \\
		\hline
		
		\tabincell{c}{Kinematic viscosity\\$\upsilon$ (${10^{ - 7}}{{\rm{m}}^2} \cdot {{\rm{s}}^{ - 1}}$)}  & & & 18.534 & & 11.887 & & 8.420 & & \\
		\hline
		
		\tabincell{c}{Salinity diffusivity\\$\alpha _{\rm{S}}$ (${10^{ - 10}}{{\rm{m}}^2} \cdot {{\rm{s}}^{ - 1}}$)}  & & & 7.75 & & 12.86 & & 18.46 & & \\
		\hline
		
		\tabincell{c}{Temperature Prandtl number\\ ${Pr _{\rm{T}}}$ (Dimensionless)}  & & & 13.349 & & 8.205 & & 5.596 & & \\
		\hline
		
		\tabincell{c}{Salinity Schmidt number\\ ${Pr _{\rm{S}}}$ (Dimensionless)}  & & & 2393.2 & & 924.3 & & 456.1 & & \\
		\hline
		
		\tabincell{c}{Kolmogorov microscale $\eta $ ($10^{-5}\rm{m}$)\\when $\varepsilon  = 1 \times {10^{ - 2}}{{\rm{m}}^2}{{\rm{s}}^{ - 3}}$} & & & 15.885 & & 11.384 & & 8.790 & & \\
		\hline
	\end{tabular}
	\label{tab3}
\end{table}

\section*{Acknowledgment}
The authors thank Mohammed Elamassie, Ming-Yuan Ren and Jian-Dong Zhang for their helpful comments, Wei-Qiang Ding for interesting discussions on numerical methods, and Andreas Muschinski for providing several manuscript documents.
Thanks are also due to John Lienhard and his co-workers for developing the website http://web.mit.edu/seawater .
We also thank Han-Tao Wang, Ying Sun, Zeng-Quan Yan, Han-Bing Wang and Fan Zhang for their early helpful work in this topic.

\section*{References}

\end{document}